\def\year{2015}
\documentclass[letterpaper]{article}
\usepackage[usenames,dvipsnames]{color}
\usepackage[T1]{fontenc}
\usepackage{aaai}
\usepackage{times}
\usepackage{helvet}
\usepackage{courier}
\usepackage{url}
\usepackage{verbatim} 
\usepackage{xspace}
\usepackage{graphicx}
\usepackage{epsfig}
\usepackage{amsmath}
\usepackage{amssymb}
\usepackage{enumitem}
\usepackage{multirow}
\usepackage{booktabs}
\newcommand{\ra}[1]{\renewcommand{\arraystretch}{#1}}
\usepackage{caption}
\usepackage{subcaption}
\usepackage{tabularx}
\usepackage{enumitem}

\newcommand{\hide}[1]{}
\newcommand{\xhdr}[1]{\vspace{1.7mm}\noindent{{\bf #1.}}}
\newcommand{\xhdrr}[1]{\vspace{1.7mm}\noindent{{\bf #1}}}

\newcommand{\eg}{{e.g.,}\xspace}
\newcommand{\ie}{{i.e.,}\xspace}

\newcommand\deem{\textcolor{Gray}}

\graphicspath{{./FIG/}}

\begin{document}

\title{Antisocial Behavior in Online Discussion Communities}
\author{Justin Cheng$^*$, Cristian~Danescu-Niculescu-Mizil$^\dagger$, Jure Leskovec$^*$\\
$^*$Stanford University, $^\dagger$Cornell University
}
\maketitle

\begin{abstract}

User contributions in the form of posts, comments, and votes are essential to the success of online communities.
However, allowing user participation also invites undesirable behavior such as trolling.
In this paper, we characterize antisocial behavior in three large online discussion communities by analyzing users who were banned from these communities.
We find that such users tend to concentrate their efforts in a small number of threads, are more likely to post irrelevantly, and are more successful at garnering responses from other users.
Studying the evolution of these users from the moment they join a community up to when they get banned, we find that not only do they write worse than other users over time, but they also become increasingly less tolerated by the community.
Further, we discover that antisocial behavior is exacerbated when community feedback is overly harsh.
Our analysis also reveals distinct groups of users with different levels of antisocial behavior that can change over time.
We use these insights to identify antisocial users early on, a task of high practical importance to community maintainers.

\end{abstract}

\section{Introduction}
\label{sec:intro}

User-generated content is critical to the success of any online platforms.
On news websites such as CNN, users comment on articles and rate the comments of other users; on social networks such as Facebook, users contribute posts that others can then comment and vote on; on Q\&A communities such as StackOverflow, users contribute and rate questions and answers.
These sites engage their users by allowing them to contribute and discuss content, strengthening their sense of ownership and loyalty~\cite{binns2012don}.

While most users tend to be civil, others may engage in antisocial behavior, negatively affecting other users and harming the community.
Such undesired behavior, which includes trolling, flaming, bullying, and harassment, is exacerbated by the fact that people tend to be less inhibited in their online interactions~\cite{suler2004online}.

Many platforms implement mechanisms designed to discourage antisocial behavior.
These include community moderation, up- and down-voting, the ability to report posts, mute functionality, and more drastically, completely blocking users' ability to post.
Additionally, algorithmic ranking attempts to hide undesirable content~\cite{hsu2009ranking}.
Still, antisocial behavior is a significant problem that can result in offline harassment and threats of violence \cite{wiener1998negligent}.

Despite its severity and prevalence, surprisingly little is known about online antisocial behavior.
While some work has tried to experimentally establish causal links, for example, between personality type and trolling \cite{buckels2014trolls}, most research reports qualitative studies that focus on characterizing antisocial behavior~\cite{donath1999identity,hardaker2010trolling}, often by studying the behavior of a small number of users in specific communities~\cite{Herring:TheInformationSociety:2011,Shachaf:JournalOfInformationScience:2010}.
A more complete understanding of antisocial behavior requires a quantitative, large-scale, longitudinal analysis of this phenomenon.
This can lead to new methods for identifying undesirable users and minimizing troll-like behavior, which can ultimately result in healthier online communities.

\xhdr{The present work}
In this paper, we characterize forms of antisocial behavior in large online discussion communities.
We use retrospective longitudinal analyses to quantify such behavior throughout an individual user's tenure in a community.
This enables us to address several questions about antisocial behavior:
First, are there users that only become antisocial later in their community life, or is deviant behavior innate?
Second, does a community's reaction to users' antisocial behavior help them improve, or does it instead cause them to become more antisocial?
Last, can antisocial users be effectively identified early on?

To answer these questions, we examine three large online discussion-based communities: {\em CNN.com}, a general news site, {\em Breitbart.com}, a political news site, and {\em IGN.com}, a computer gaming site.
On these sites, editors and journalists post articles on which users can then comment.
We study complete data from these websites: over 18 months, 1.7 million users contributed nearly 40 million posts and more than 100 million votes.
In these communities, members that repeatedly violate community norms are eventually banned permanently.
Such individuals are clear instances of antisocial users, and constitute ``ground truth'' in our analyses.

\xhdr{Characterizing antisocial behavior}
We compare the activity of users who are later banned from a community, or \emph{Future-Banned Users (FBUs)}, with that of users who were never banned, or \emph{Never-Banned Users (NBUs)}.
By analyzing the language of their posts, we find significant differences between these two groups.
For example, FBUs tend to write less similarly to other users, and their posts are harder to understand according to standard readability metrics.
They are also more likely to use language that may stir further conflict (e.g., they use less positive words and use more profanity).

FBUs also differ in how they engage in discussions: their posts tend to be concentrated in individual threads rather than spread out across several. 
They receive more replies than average users, suggesting that they might be successful in luring others into fruitless, time-consuming discussions.

\xhdr{Longitudinal analysis}
We find that the behavior of an FBU worsens over their active tenure in a community.
Through a combination of crowdsourcing experiments and machine learning, we show that not only do they enter a community writing worse posts than NBUs, but the quality of their posts also worsens more over time.  This suggests that communities may play a part in incubating antisocial behavior.
In fact, users who are excessively censored early in their lives are more likely to exhibit antisocial behavior later on.
Furthermore, while communities appear initially forgiving (and are relatively slow to ban these antisocial users), they become less tolerant of such users the longer they remain in a community.
This results in an increased rate at which their posts are deleted, even after controlling for post quality.

\xhdr{Typology of antisocial users}
Among FBUs, we observe that the distribution of users' post deletion rates (i.e., the proportion of a user's posts that get deleted by moderators) is bimodal.
Some FBUs have high post deletion rates, while others have relatively low deletion rates.
While both types of FBUs tend to write similarly overall, those with high post deletion rates write less similarly to other users in the same discussion thread and write more in each discussion they participate in, while those with low post deletion rates spread their posts across a larger number of discussions, and thus attract less attention.
Starting from this observation, we introduce a typology of antisocial behavior based on comparing a user's post deletion rate across the first and second halves of their life, and identify users who are getting worse over time, as well as those who later redeem themselves.

\xhdr{Predicting future banning}
Last, we show that a user's posting behavior can be used to make predictions about who will be banned in the future.
Inspired by our empirical analysis, we design features that capture various aspects of antisocial behavior: post content, user activity, community response, and the actions of community moderators.
We find that we can predict with over 80\% AUC (area under the ROC curve) whether a user will be subsequently banned.
In fact, we only need to observe 5 to 10 user's posts before a classifier is able to make a reliable prediction.
Further, cross-domain classification performance remains high, suggesting that the features indicative of antisocial behavior that we discover are not community-specific.

Antisocial behavior is an increasingly severe problem that currently requires large amounts of manual labor to tame.
Our methods can effectively identify antisocial users early in their community lives and alleviate some of this burden.

\section{Related Work}
\label{sec:related}

\begin{table*}[t]
\small
\centering
\ra{1.3}
\begin{tabular*}{\textwidth}{@{\extracolsep{\fill}}lrrrrrr}\toprule
\textbf{Community} & \textbf{\# Users} & \textbf{\# Users Banned} & \textbf{\# Threads} & \textbf{\# Posts} & \textbf{\# Posts Deleted} & \textbf{\# Posts Reported} \\
\hline
CNN & 1,158,947 & 37,627 \deem{(3.3\%)} & 200,576 & 26,552,104 & 5,355,344 \deem{(21.4\%)} & 1,156,005 \deem{(4.6\%)} \\
IGN & 343,926 & 5,706 \deem{(1.7\%)} & 682,870 & 7,967,414 & 184,643 \deem{(2.3\%)} & 88,621 \deem{(1.1\%)} \\
Breitbart & 246,422 & 5,350 \deem{(2.2\%)} & 376,526 & 4,376,369 & 119,265 \deem{(2.7\%)} & 117,779 \deem{(2.7\%)} \\
\bottomrule
\end{tabular*}
\caption{Summary statistics of the three large news discussion communities analyzed. Percentages of totals are in 
parentheses.
}
\label{tab:dataset}
\end{table*}

We start by considering definitions of antisocial behavior, summarize work on antisocial behavior online, then discuss how such behavior can be detected in a variety of settings.

\xhdr{Antisocial behavior}
Antisocial behavior, which includes trolling, flaming, and griefing, has been widely discussed in past literature.
For instance, a troll has been defined as a person that engages in ``negatively marked online behavior'' \cite{hardaker2010trolling}, or a user who initially pretends to be a legitimate participant but later attempts to disrupt the community \cite{donath1999identity}.
Trolls have also been characterized as ``creatures who take pleasure in upsetting others'' \cite{kirman2012exploring}, and indeed, recent work has found that sadism is strongly associated with trolling tendencies \cite{buckels2014trolls}.
Finally, some literature instead provides a taxonomy of deviant behavior \cite{suler1998bad}.
In this paper, we rely on a community and its moderators to decide who they consider to be disruptive and harmful, and conduct aggregate analyses of users who were permanently banned from a community.

\xhdr{Studying antisocial behavior}
Research around antisocial behavior has tended to be largely qualitative, generally involving deep case study analyses of a small number of manually-identified trolls. 
These analyses include the different types of trolling that occur \cite{hardaker2010trolling}, the motivations behind doing so \cite{Shachaf:JournalOfInformationScience:2010}, and the different strategies that others use in response to trolls \cite{baker2001moral,chesney2009griefing}. 
Other work has quantified the extent of such negative behavior online \cite{juvonen2008extending}.
In contrast, the present work presents a large-scale data-driven analysis of antisocial behavior in three large online communities, with the goal of obtaining quantitative insights and developing tools for the early detection of trolls.
Conceptually related is a prior study of the effects of community feedback on user behavior~\cite{cheng2014community}, which revealed that negative feedback can lead to antisocial behavior.
However, rather than focusing on individual posts, this paper takes a longer-term approach and studies antisocial users and their evolution throughout their community life.

\xhdr{Detecting antisocial behavior}
Several papers have focused on detecting vandalism on Wikipedia by using features such as user language and reputation, as well as article metadata~\cite{adler2011wikipedia,potthast2008automatic}.
Other work has identified undesirable comments based their relevance to the discussed article and the presence of insults~\cite{sood2012automatic},
and predicted whether players in an online game would be subsequently punished for reported instances of bad behavior~\cite{blackburn2014stfu}.
Rather than predicting whether a particular edit or comment is malicious, or focusing only on cases of bad behavior, we instead predict whether individual users will be subsequently banned from a community based on their overall activity, and show how our models generalize across multiple communities.
Nonetheless, the text and post-based features used in this prediction task are partially inspired by those used in prior work.

\section{Data Preparation}
\label{sec:data}

\xhdr{Dataset description}
We investigated three online news communities: \emph{CNN.com} (general news), \emph{Breitbart.com} (political news), and \emph{IGN.com} (computer gaming), selected based on their large size (Table \ref{tab:dataset}).
On these sites, community members post comments on (news) articles, and each comment can either be replied to, or voted on.
In this paper, we refer to comments and replies as \emph{posts}, and to the list of posts on the same article as a \emph{thread}.
Disqus, a commenting platform that hosts the discussions in these communities, provided us with a complete timestamped trace of user activity from March 2012 to August 2013, as well as a list of users that were banned from posting in these communities.

\xhdr{Measuring undesired behavior}
On a discussion forum, undesirable behavior may be signaled in several ways: users may \emph{down-vote}, \emph{comment} on, or \emph{report} a post, and community moderators may \emph{delete} the offending post or outright \emph{ban} a user from ever posting again in the forum.
However, down-voting may signal disagreement rather than undesirability.
Also, many web sites such as Breitbart have low down-voting rates (only 4\% of all votes are down-votes); others may simply not allow for down-voting.
Further, one would need to define arbitrary thresholds (\eg a certain fraction of down-votes) needed to label a user as antisocial.
Automatically identifying undesirable posts based on the content of replies may also be unreliable.
In contrast, we find that post deletions are a highly precise indicator of undesirable behavior, as only community moderators can delete posts.
Moderators generally act in accordance with a community's comment policy, which typically covers disrespectfulness, discrimination, insults, profanity, or spam.
Post reports are correlated with deletions, as these reported posts are likely to be subsequently deleted.

At the user-level, bans are similarly strong indicators of antisocial behavior, as only community moderators can ban users.
Empirically, we find that many of these banned users exhibit such behavior.
Apart from insults and profanity, these include repeated attempts to bait users (``Ouch, ya got me. What's Google?''), provoke arguments (``Liberalism truly breeds violence...this is evidence of that FACT''), or derail discussions (``All I want to know is...was there a broom involved in any shape or form?'').

Thus, we focus on users who moderators have subsequently banned from a community and refer to them as \emph{Future-Banned Users} or \emph{FBUs}.
While such an approach does not necessarily identify every antisocial user, this results in a more precise set of users who were explicitly labeled as undesirable.
Further, we restrict our analysis to banned users with at least five posts who joined a given community after March 2012, so that we can track their behavior from the beginning of their community life to the time they are banned.
We excluded users who were banned multiple times so as not to confound the effects of temporary bans with behavior change, as well as users whose posts contained URLs to remove link spammers.
After filtering, we obtained a core set of banned users for each community ($N$=10,476 for CNN, 660 for IGN, and 736 for Breitbart) for which we had a complete trace of activity.

\xhdr{Matching FBUs and NBUs} First, we note that FBUs tend to post more frequently than average users (prior to getting banned): on CNN, a typical FBU makes 264 posts, but an average user makes only 22 posts.
To control for this large disparity in posting activity, we use matching \cite{rosenbaum1983central}, a statistical technique used to support causality claims in observational studies, to control for the number of posts a user made and the number of posts made per day.
In other words, for each FBU, or user who was later banned, we identify a similarly active user that was never banned (a \emph{Never-Banned User} or \emph{NBU}).

\begin{figure*}[tb]
        \centering
        \begin{subfigure}[b]{0.32\textwidth}
                \includegraphics[width=\textwidth]{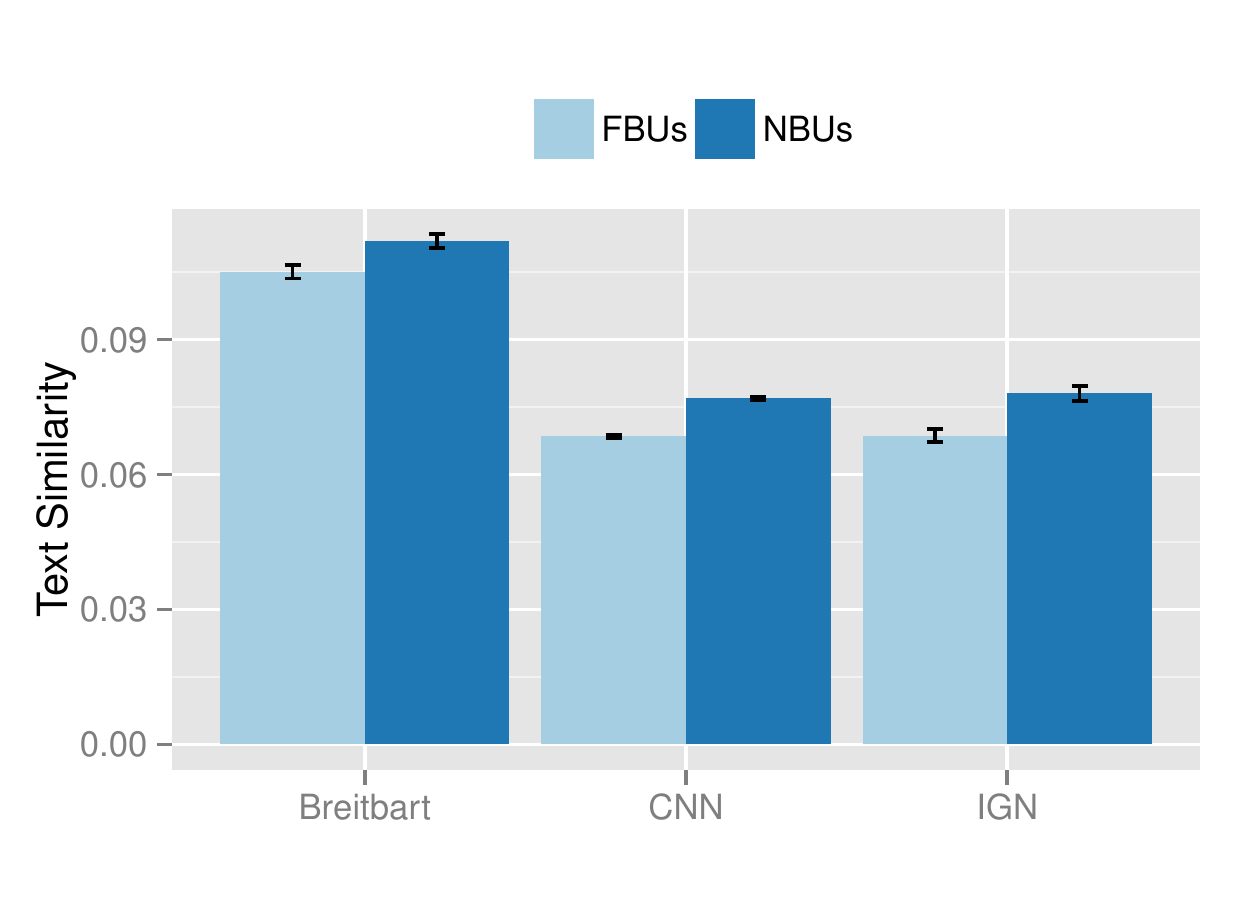}
                \caption{Text Sim. of a Post with its Parent Thread}
                \label{fig:plot_cosine}
        \end{subfigure}
        \begin{subfigure}[b]{0.32\textwidth}
                \includegraphics[width=\textwidth]{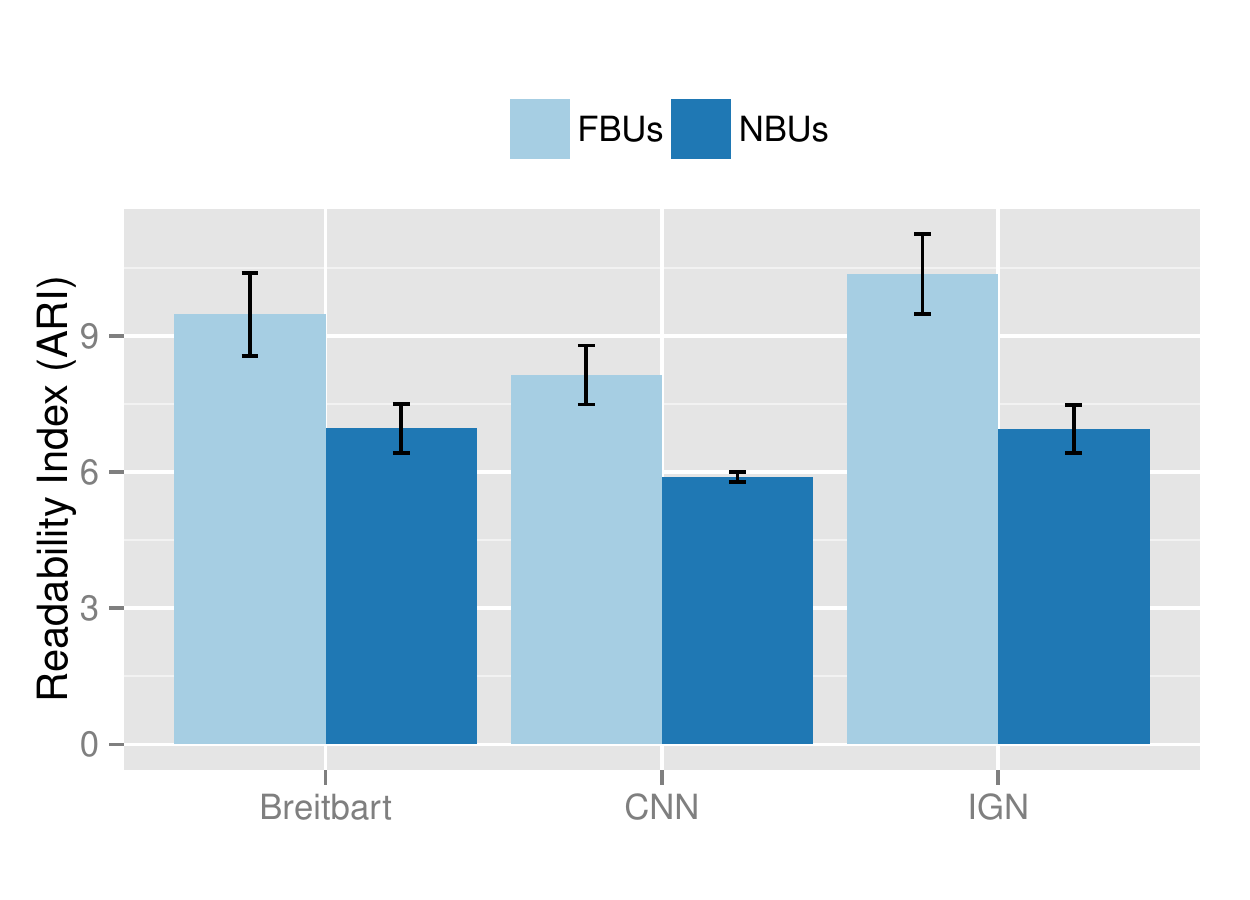}
                \caption{Readability Index}
                \label{fig:plot_readability}
        \end{subfigure}
        \begin{subfigure}[b]{0.32\textwidth}
                \includegraphics[width=\textwidth]{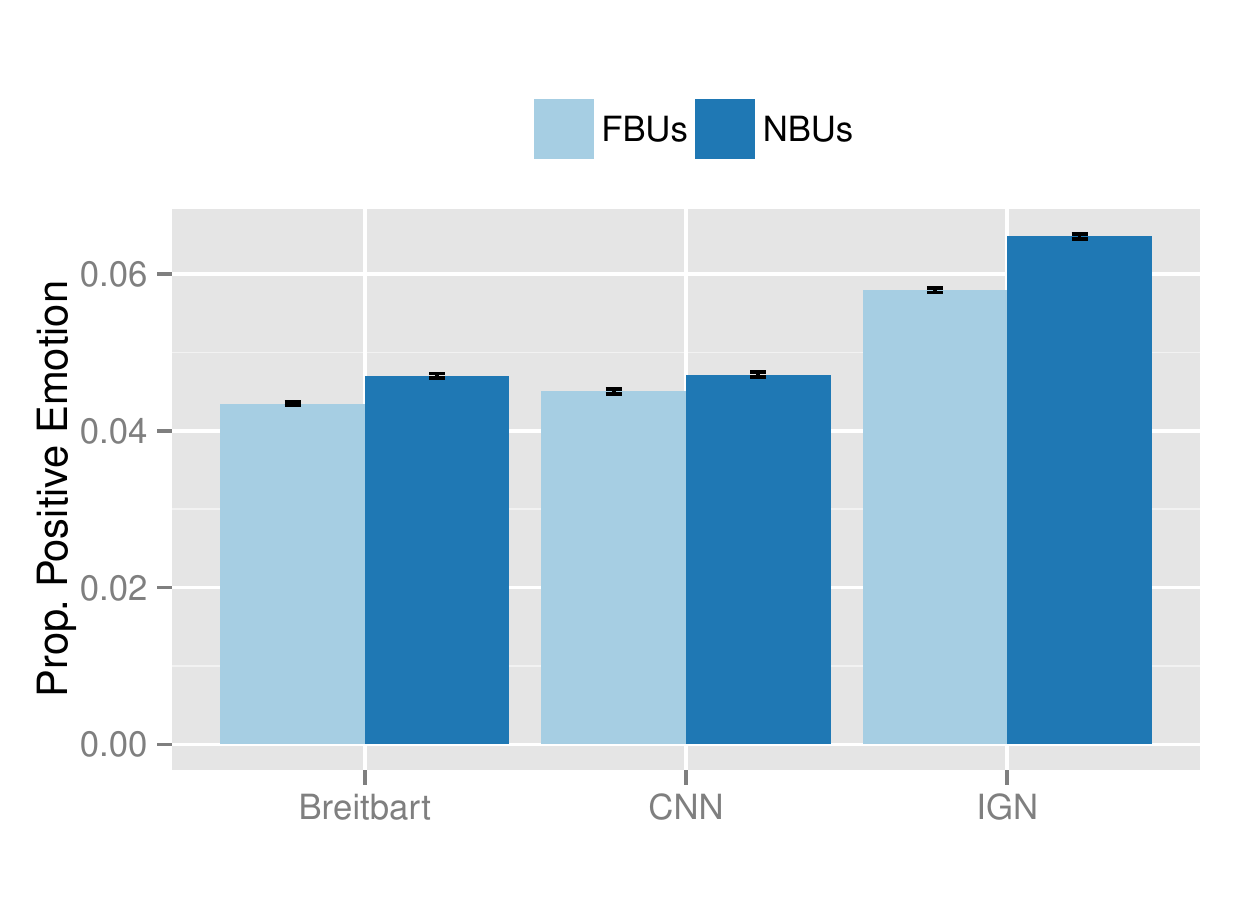}
                \caption{Positive Emotion}
                \label{fig:plot_posemo}
        \end{subfigure}%
        \caption{Users who get banned in the future (FBUs) (a) write less similarly to other users in the same thread, (b) write posts that are harder to read 
        (i.e., have a higher readability index),
         and (c) express less positive emotion.}\label{fig:writing_behavior}
\end{figure*}

\xhdr{Measuring text quality}
How might we obtain an unbiased measure of the quality or appropriateness of a post?
Dictionary-based approaches may miss non-dictionary words (e.g., ``Obummer''); a classifier trained on the text of deleted posts may confound post content and community bias, as communities tend to develop animosity towards undesirable users over time and over-penalize them \cite{cheng2014community}.
Thus, we instead obtained human judgments of the appropriateness of a post, and collected labels for a random sample of 6000 posts using Mechanical Turk.
To construct this dataset, 500 FBUs and NBUs were sampled from each of the three communities, and two posts sampled from each user.
Using the text of a post alone, workers were asked, on a scale of 1 to 5, to evaluate how appropriate a post was given general moderation guidelines (\ie to look out for disrespectfulness, discrimination, insults, profanity, or spam).
Each post was labeled by three independent workers, and their ratings averaged (Krippendorff's $\alpha$=0.39).
131 workers completed these tasks, and they rated deleted posts significantly lower than non-deleted posts (2.4 vs. 3.0, \textit{p}<10\textsuperscript{-4}).

Using these labeled posts, we then trained a logistic regression model on text bigrams to predict the text quality (or appropriateness) of a post.
Posts with a rating higher than 3 were labeled as appropriate, and those a rating of 3 or lower were labeled as inappropriate.
Under ten-fold cross validation, the AUC attained by this classifier was 0.70.
This suggests that while the classifier is able to partially capture this human decision making process and allow us to observe overall trends in the data, other factors may play a significant role in determining whether a post is appropriate.

Finally, the results reported in this paper apply to all three communities studied unless otherwise noted.
All figures are plotted using data from the CNN community, and the figures for other communities are qualitatively similar.
Error bars represent the standard error of the mean.

\section{Understanding Antisocial Behavior}
\label{sec:behavior}

To understand antisocial behavior in the context of a discussion community, we first characterize how users who are subsequently banned differ from those who are not in terms of how they write, as well as how they act in a community.
Then, we analyze changes in behavior over the lifetimes of these users to understand the effects of post quality, community bias, and excessive censorship.

\subsection{Characterizing Antisocial Behavior}

Corroborating previous literature, we find significant differences in how FBUs and NBUs write, even after controlling for similar posting activity (e.g., the number of posts made).

\xhdrr{How do FBUs write?}
The similarity of a post to previous posts in a same thread may reveal how users are contributing to a community.
Users can stay on-topic or veer off-topic; prior work has also shown that users tend to adopt linguistic conventions or jargon in a community~\cite{danescu2013no} and that they also unconsciously mimic the choices of function-word classes they are communicating with~\cite{DanescuNiculescuMizil:ProceedingsOfWww:2011a}.
Here, we compare the average text similarity of a user's post with the previous three posts in the same thread (if they exist), obtained by computing the cosine similarity of words used in these posts.
We find that the average text similarity of posts written by FBUs is significantly lower than that of NBUs (\textit{t}(20950)=16, \textit{p}<10\textsuperscript{-4}, Cohen's \textit{d}=0.25 for CNN, \textit{t}(1318)=4.2, \textit{p}<10\textsuperscript{-4}, \textit{d}=0.30 for IGN, and \textit{t}(1470)=3.1, \textit{p}<0.01, \textit{d}=0.20 for Breitbart) (Figure \ref{fig:plot_cosine}), suggesting that FBUs make less of an effort to integrate or stay on-topic.
Overall, we find that post deletion is weakly negatively correlated with text similarity ($r$=-0.08), which suggests that off-topic posts are more likely to be deleted.

Next, we measure each post with respect to several readability tests, including the Automated Readability Index (ARI), which are designed to gauge how understandable a piece of text is. While posts written by FBUs and NBUs have similar word counts, those written by FBUs have a higher ARI, and thus appear to be less readable than those written by NBUs (\textit{t}>2.36, \textit{p}<0.01, \textit{d}>0.05) (Figure \ref{fig:plot_readability}).

Prior research also suggests that trolls tend to make inflammatory posts \cite{hardaker2010trolling}, and these observations hold true here for FBUs and NBUs, though these effects are relatively small.
We compare the proportion of words used in different LIWC categories \cite{pennebaker2001linguistic} to identify the different aspects in which the language used by FBUs and NBUs differ (but nonetheless note that this approach may miss non-dictionary words).
While there was no consistent trend with respect to the use of words connotating negative emotion, we find that FBUs are less likely to use positive words (\textit{t}>4.58, \textit{p}<10\textsuperscript{-4}, \textit{d}>0.06) (Figure \ref{fig:plot_posemo}), corroborating research that suggests positive replies to comments minimize conflict \cite{laniado2012emotions}.
FBUs are also more likely to swear (\textit{t}>2.70, \textit{p}<0.01, \textit{d}>0.05), or use less tentative or conciliatory language (\ie~less use of words such as ``could'', ``perhaps'', or ``consider'') (\textit{t}>2.50, \textit{p}<0.05, \textit{d}>0.08).

\begin{figure*}[ht]
        \centering
        \begin{subfigure}[b]{0.32\textwidth}
                \includegraphics[width=\textwidth]{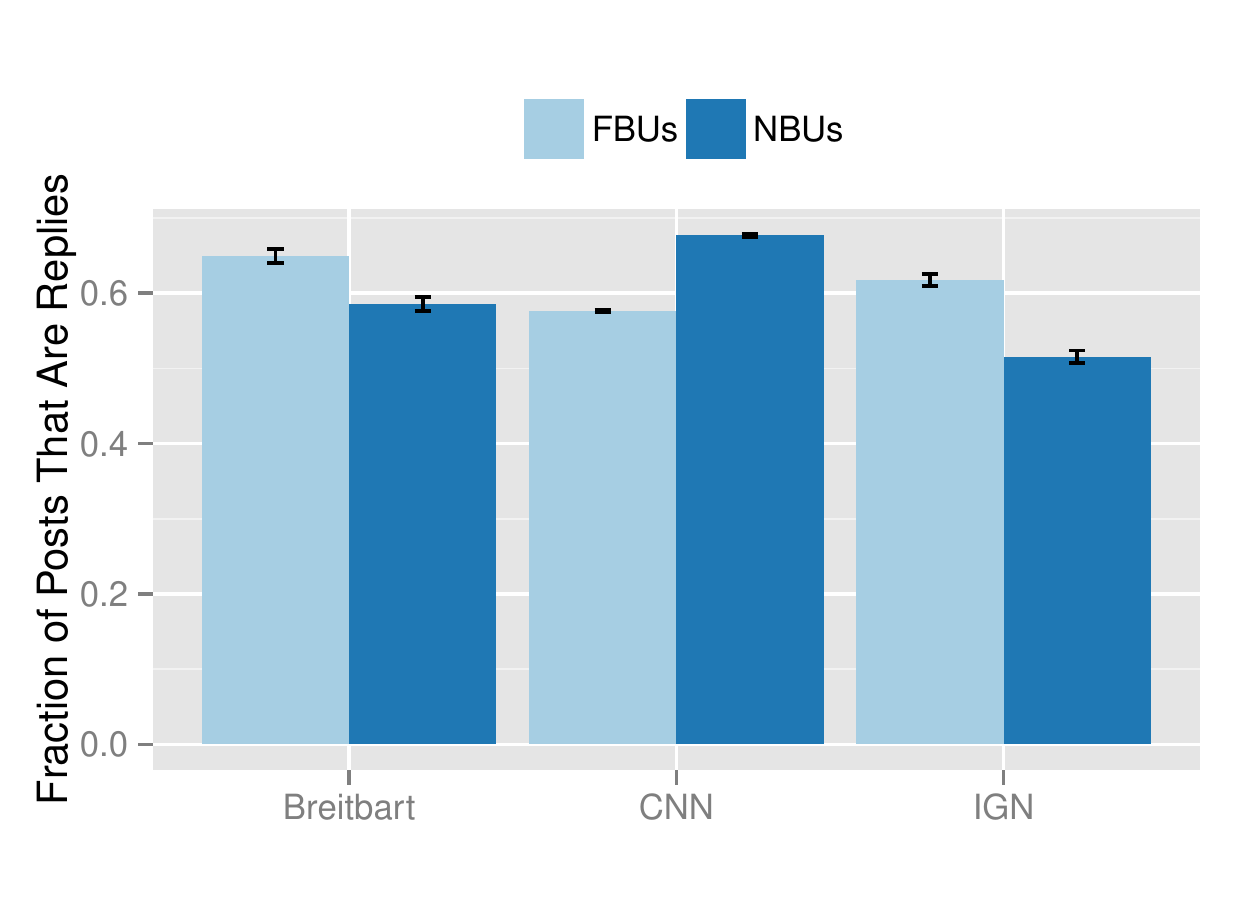}
                \caption{Fraction of Posts That Are Replies}
                \label{fig:plot_reply}
        \end{subfigure}
        \begin{subfigure}[b]{0.32\textwidth}
                \includegraphics[width=\textwidth]{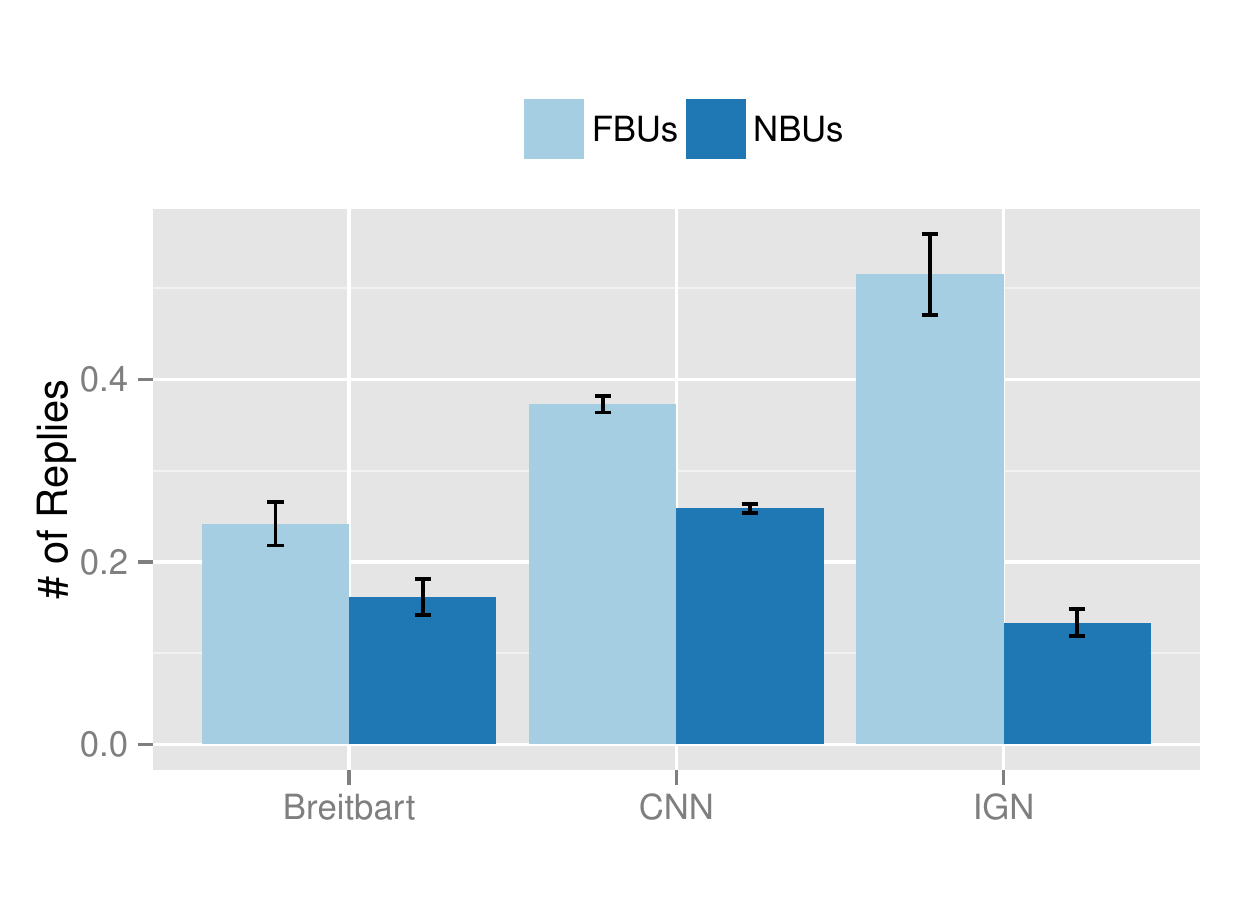}
                \caption{\# Replies}
                \label{fig:plot_replies}
        \end{subfigure}
        \begin{subfigure}[b]{0.32\textwidth}
                \includegraphics[width=\textwidth]{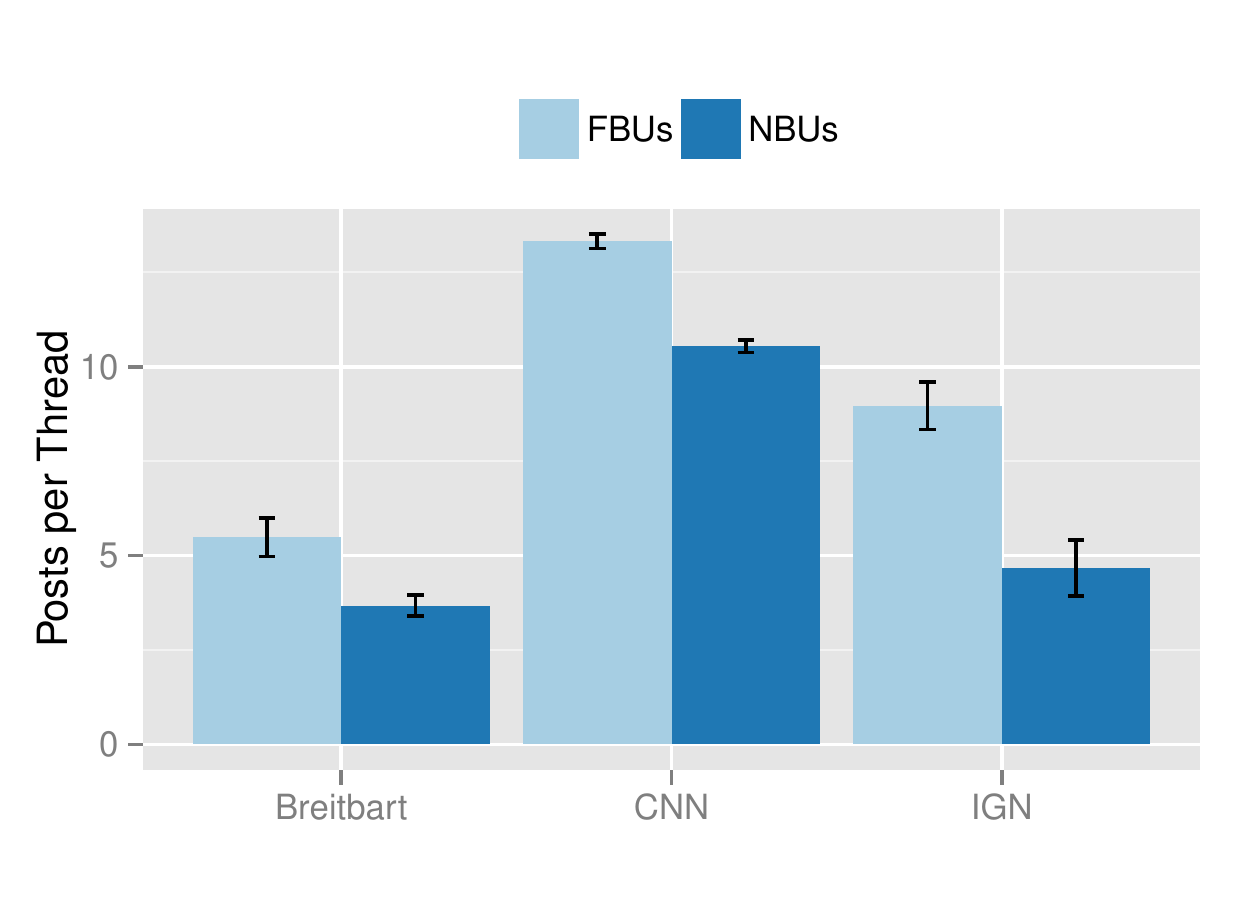}
                \caption{Posts per Thread}
                \label{fig:plot_posts_per_thread}
        \end{subfigure}
        \caption{While FBUs may either (a) create or contribute to (already existing) discussions depending on the communty, they generally (b) get more replies from other users, and (c) 
        concentrate on fewer threads.}\label{fig:activity_behavior}
\end{figure*}

\xhdrr{How do FBUs generate activity around themselves?}
Do FBUs purposefully try to create discussions, or opportunistically respond to an on-going discussion?
Here, we find that their behavior differs depending on the community (Figure \ref{fig:plot_reply}).
On Breitbart and IGN, trolls are more likely to reply to others' posts, but on CNN, they are more likely to start new discussions (\textit{t}>9.1, \textit{p}<10\textsuperscript{-4}, \textit{d}>0.25).
Still, across all communities, FBUs appear to be effective at luring other users into potentially fruitless discussions, supporting claims in previous literature about troll-like behavior \cite{Herring:TheInformationSociety:2011}: the average number of replies a FBU gets is significantly higher than that of a regular user (\textit{t}>2.6, \textit{p}<0.01, \textit{d}>0.16, Figure \ref{fig:plot_replies}).
We observe a similar trend if we instead consider all descendant posts instead of only direct replies (to quantify the total discussion volume generated by a post).

Last, FBUs contribute significantly more posts per thread they participate in (\textit{t}>3.1, \textit{p}<0.01, \textit{d}>0.16, Figure \ref{fig:plot_posts_per_thread}), perhaps engaged in protracted discussions with other users.

\subsection{Evolution Over Time}

While FBUs behave differently from NBUs, how does their behavior and the community's perception of them change over time?
Here, we plot the post deletion rates (or proportion of posts deleted) of both types of users over time, and where each user made at least 10 posts (before being banned) ($N$=9379 for CNN, 582 for IGN, 569 for Breitbart).
On CNN, these FBUs survived an average of 42 days before being banned; on IGN, they survived 103 days; on Breitbart, 82 days.
Time, as defined by the index of a post, was normalized across users.
Figure \ref{fig:deleted_prob} shows that on average, the deletion rate of an FBU's posts tends to increase over their life in a community.
In contrast, the deletion rate for NBUs remains relatively constant.

The increase in the post deletion rate could have two causes: (H1) a decrease in posting quality --- that FBUs tend to write worse later in their life;
 or, (H2) an increase in community bias --- that the community starts to recognize these users over time and becomes less tolerant of their behavior, thus penalizing them more heavily.

To create a measure of post quality, we predict the text quality of a post using a classifier trained on human labels obtained via crowdsourcing, as described previously.
Figure \ref{fig:deleted_quality} shows that for FBUs (and maybe NBUs), text quality is decreasing over time, suggesting that H1 may be supported.

To test both H1 and H2, we conducted a series of studies:

\begin{figure}[tb]
        \centering
        \begin{subfigure}[b]{0.48\columnwidth}
                \includegraphics[width=\textwidth]{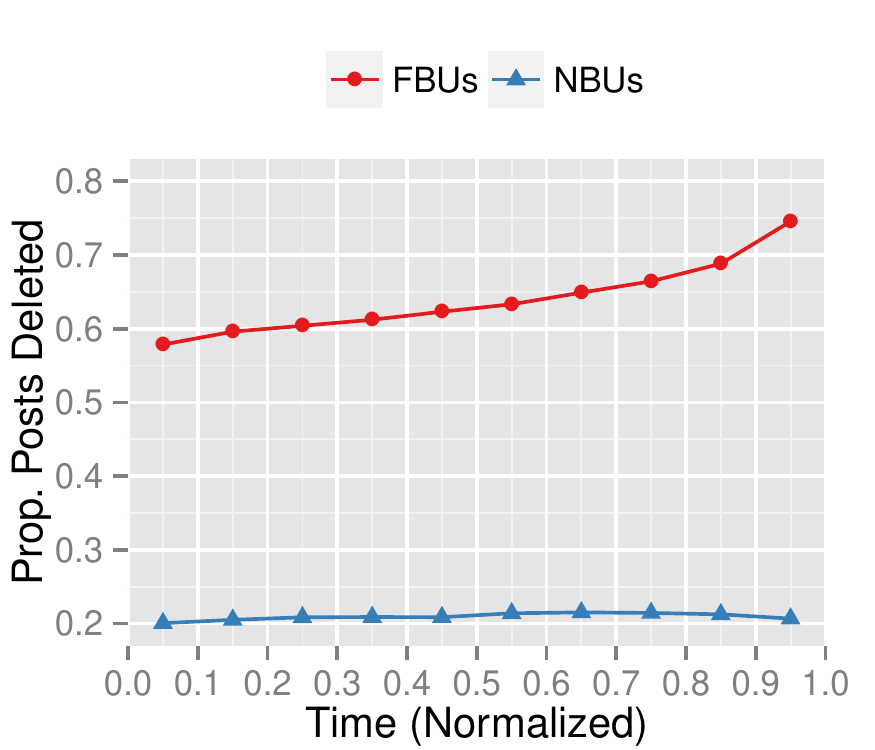}
                \caption{Post deletion rate}
                \label{fig:deleted_prob}
        \end{subfigure}
        \begin{subfigure}[b]{0.48\columnwidth}
                \includegraphics[width=\textwidth]{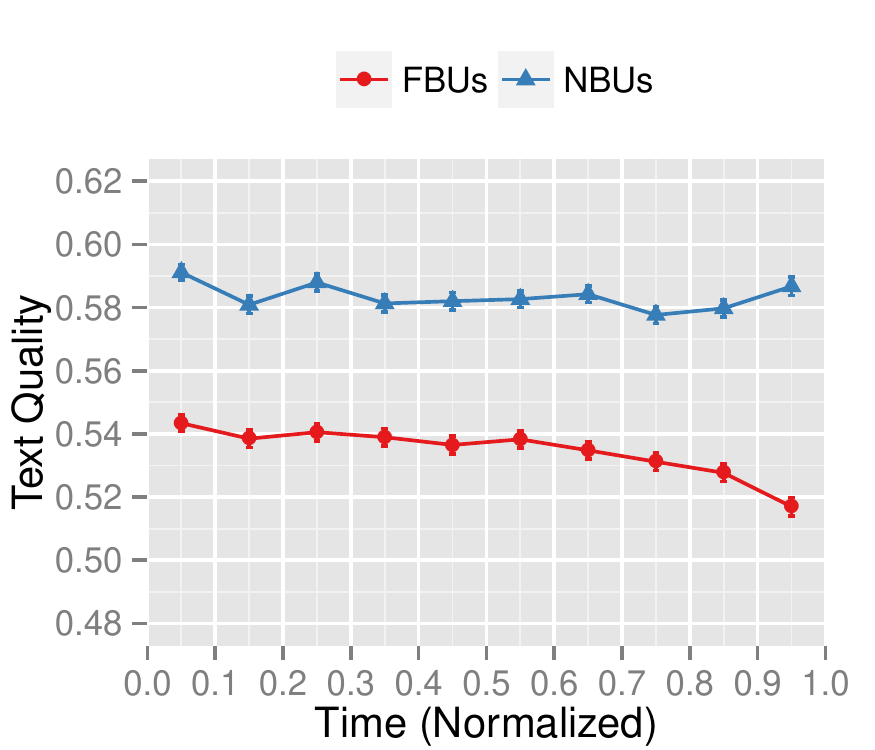}
                \caption{Text quality}
                \label{fig:deleted_quality}
        \end{subfigure}
        \caption{(a) The rate of post deletion increases over time for FBUs, but is effectively constant for NBUs. (b) Similarly, text quality decreases over time for FBUs, but not for NBUs.}
        \label{fig:deleted}
\end{figure}

\begin{table}[tb]
\small
\centering
\ra{1.3}
\begin{tabularx}{\columnwidth}{XXXX}\toprule
  & \multicolumn{3}{c}{Mean Post Appropriateness on CNN (1-5)} \\
  & \textbf{All Posts} & \textbf{First 10\%} & \textbf{Last 10\%} \\
  \hline
  FBUs & 2.7 & 3.0 & 2.3 \\
  NBUs & 3.3 & 3.5 & 3.2 \\
\bottomrule
\end{tabularx}
\caption{FBUs start out writing worse than NBUs and worsen more than NBUs over the course of their life. Higher appropriateness ratings correspond to higher quality posts.}
\label{tab:crowdquality}
\end{table}

\subsubsection{Do users write worse over time?}
While the text quality of posts by FBUs seems to be decreasing over time, is this effect significant?
To test the hypothesis that FBUs (or NBUs) tend to write worse over time, we conducted an experiment to see if people could differentiate posts written at the beginning or the end of a user's life (\ie from the time they join the community to when they get banned).
From each community, we selected 200 FBUs and 200 NBUs at random, and from each user sampled a random post from the first 10\% or last 10\% of their entire posting history.
We then shuffled and presented each post to three different workers on Mechanical Turk, and asked them to rate the appropriateness of each post on a scale of 1 to 5, subsequently averaging these ratings.
95 workers completed these tasks (Krippendorff's $\alpha$=0.35).
As Table \ref{tab:crowdquality} shows, FBUs enter a community already writing worse than NBUs (3.0 vs. 3.5 for CNN, \textit{p}<0.05 for all communities).
Moreover, for both user types, post ratings also decreased with time (\textit{p}<0.05), supporting H1 and previous work that showed that users in discussion communities tend to write worse over time \cite{cheng2014community}.
In fact, post ratings decreased more for FBUs than NBUs (\textit{p}<0.05, \textit{d}<0.19 for NBUs, \textit{d}>0.29 for FBUs).
In other words, while both FBUs and NBUs write worse over time, this change in quality is larger for FBUs.

Instead comparing the predicted text quality of posts written in the first half of a user's life with that written in the second half reveals similar findings (\textit{p}<0.01), with larger effect sizes for FBUs (\textit{d}>0.15) than NBUs (\textit{d}<0.02).
As most FBUs only survive a month or two before getting banned, larger changes may be observable over longer time periods.

\begin{figure*}[t]
        \centering
        \begin{subfigure}[b]{0.32\textwidth}
                \includegraphics[width=\textwidth]{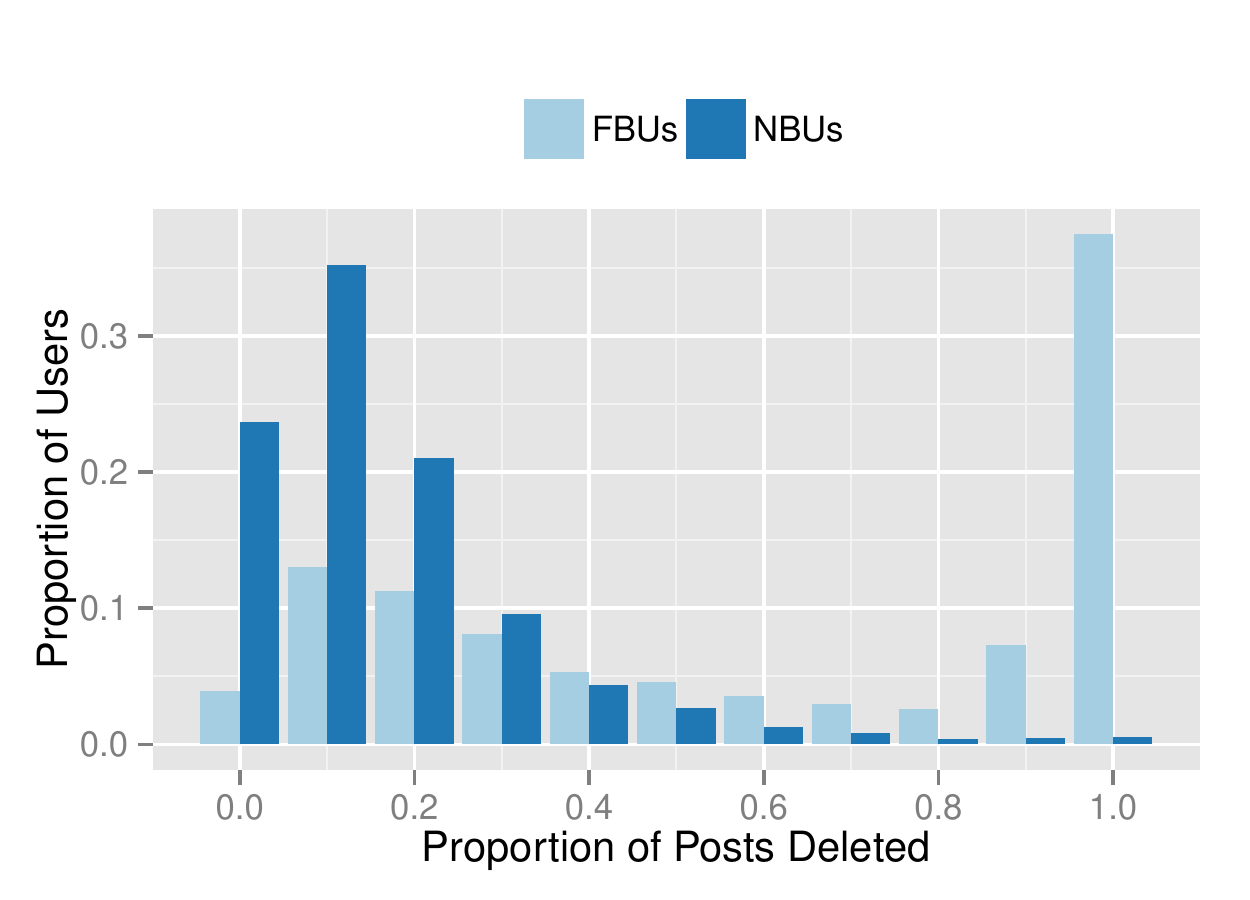}
                \caption{Distribution of Prop. Deleted Posts}
                \label{fig:plot_deletion_prop}
        \end{subfigure}
        \begin{subfigure}[b]{0.32\textwidth}
                \includegraphics[width=0.5\textwidth]{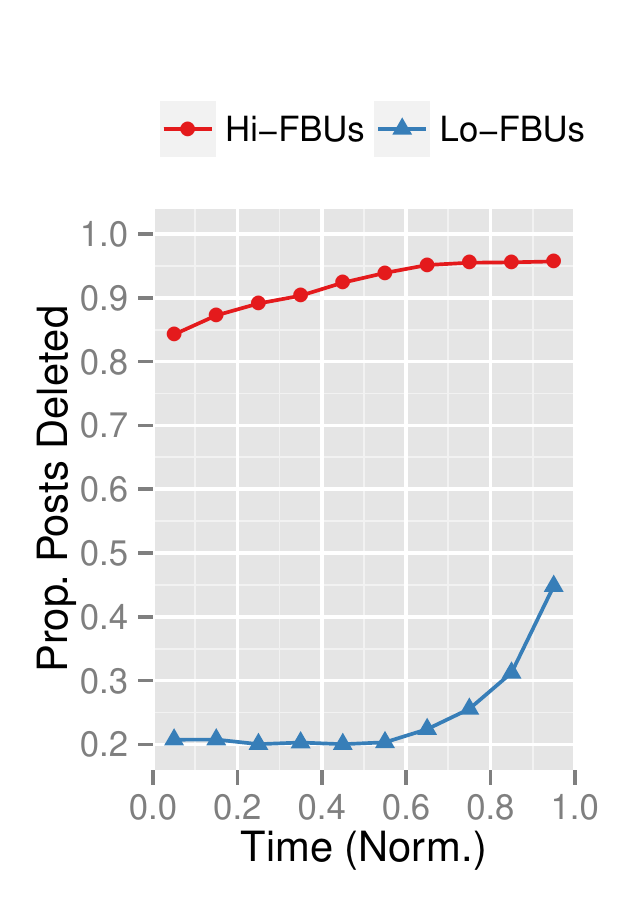}\includegraphics[width=0.5\textwidth]{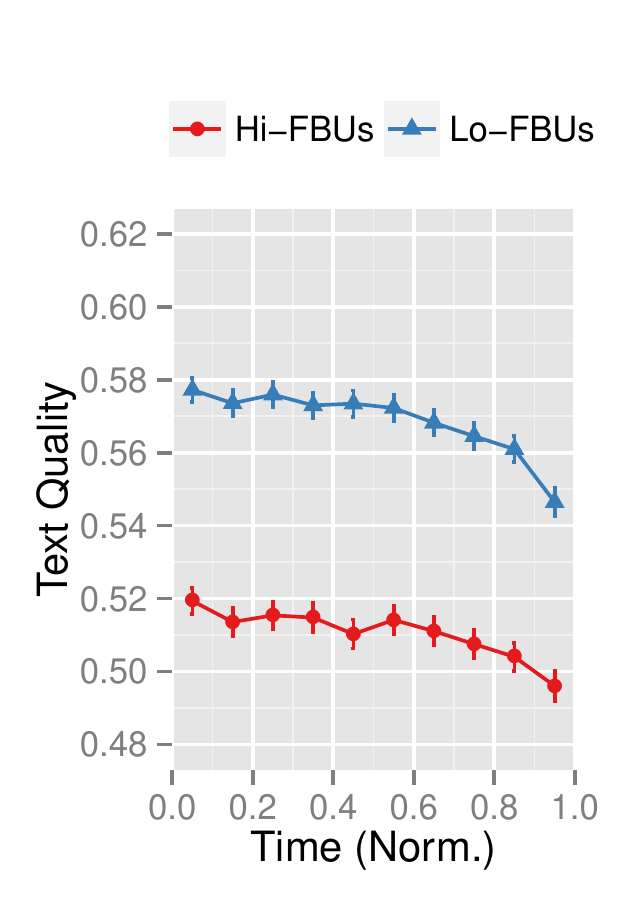}
                \caption{Persistents vs. Occasionals Over Time}
                \label{fig:plot_deleted_bad}
        \end{subfigure}
        \begin{subfigure}[b]{0.32\textwidth}
                \includegraphics[width=\textwidth]{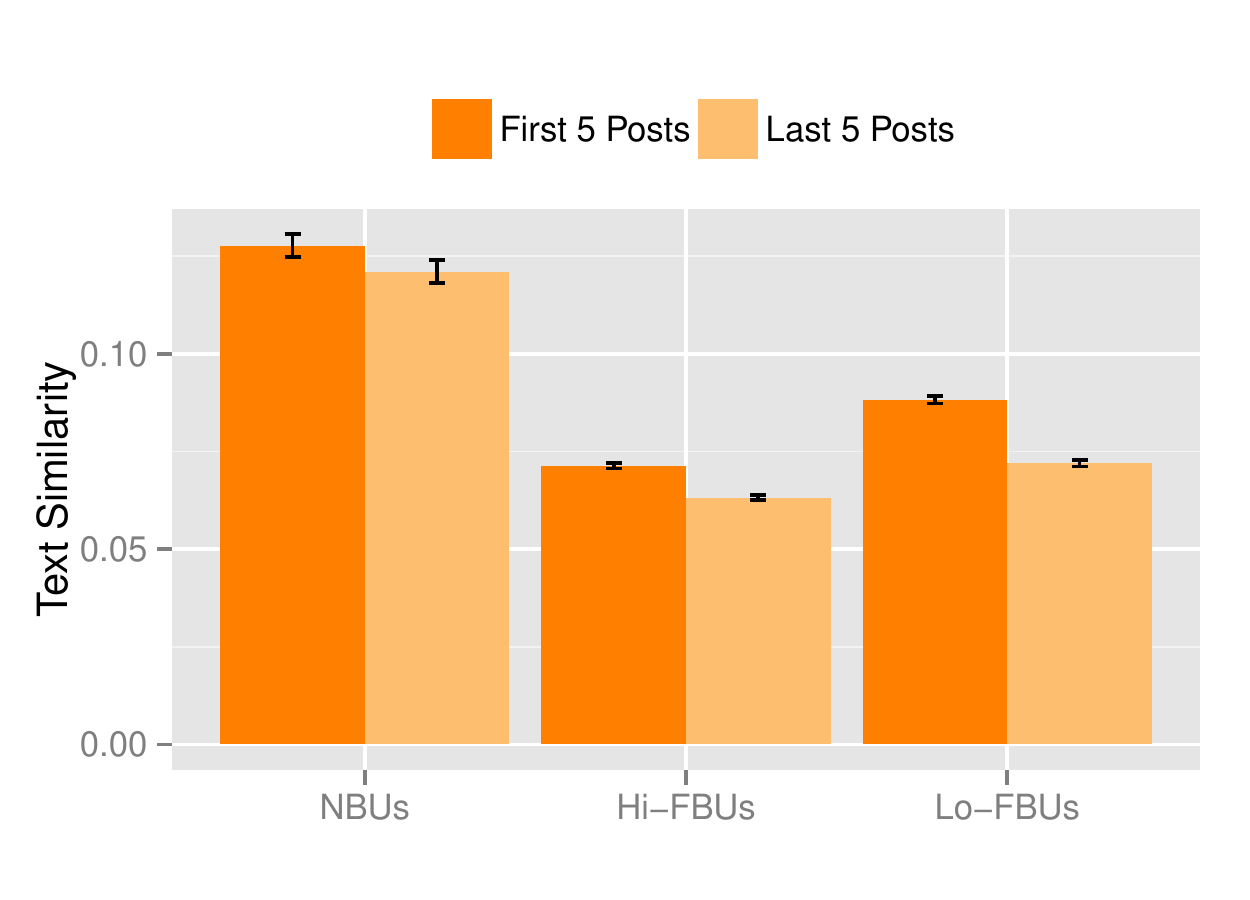}
                \caption{Thread-level Text Similarity}
                \label{fig:plot_lohiprop_cosine}
        \end{subfigure}
        \caption{(a) The distribution of FBUs' proportions of deleted posts is bimodal, suggesting that there are two types of FBUs. (b) These two groups exhibit different deletion rates over their lives, with Hi-FBUs appearing to write worse than Lo-FBUs. (c)~The change in deletion rates can be explained by a decrease in text similarity over time.}
        \label{fig:lohiprop_plots}
\end{figure*}

\subsubsection{Does community tolerance change over time?}
To test for an effect of community tolerance or bias, we used propensity score matching \cite{rosenbaum1983central} to pair random posts of similar predicted text quality.
In each pair, one post was taken from the first 10\% of a user's life, and the other taken from the final 10\% of a user's life.
After obtaining these pairs, we performed a Wilcoxon Signed-rank test to check whether a post taken from the start of a user's life is more likely than a post taken from the end of a user's life to be deleted.

Among FBUs, we find a significant effect of post time in all communities (\textit{W}>14705, \textit{p}<10\textsuperscript{-4}, effect size \textit{r}>0.21), or that a post was made in the last 10\% of an FBU's life is more likely to be deleted, supporting H2.
Among NBUs, there is no significant effect (\textit{p}>0.39, \textit{r}<0.04).
In other words, we find a community bias against FBUs: posts written later in an FBU's life are more likely to be deleted, regardless of whether they are actually worse.

\subsubsection{Does excessive censorship cause users to write worse?}
Finally, we wanted to understand if a draconian post deletion policy could exacerbate undesirable behavior.
In other words, if users had their posts arbitrarily deleted, despite writing similarly to other users whose posts were left alone, were they more likely to write worse in the future?

In this study, we considered users who wrote at least 10 posts, and computed the mean text quality of each user's first five posts.
We divided these users into two populations: those which had four or more posts deleted among their first five posts, and those who had one or less posts deleted.
Next, we matched pairs of users from these two populations on text quality, and then compare the mean text quality of their subsequent five posts.
Here, a Wilcoxon Signed-rank test shows a significant effect of the deletion rate (\textit{W}>5956, \textit{p}<0.05, \textit{r}>0.15).
Thus, given two users who initially write posts of similar quality, but where one user's posts are then ``unfairly'' deleted while the other user's posts are not, the former is more likely to write worse in the future.

\section{Types of Antisocial Users}
\label{sec:types}

\begin{figure}[tb]
    \centering
    \includegraphics[width=0.48\textwidth]{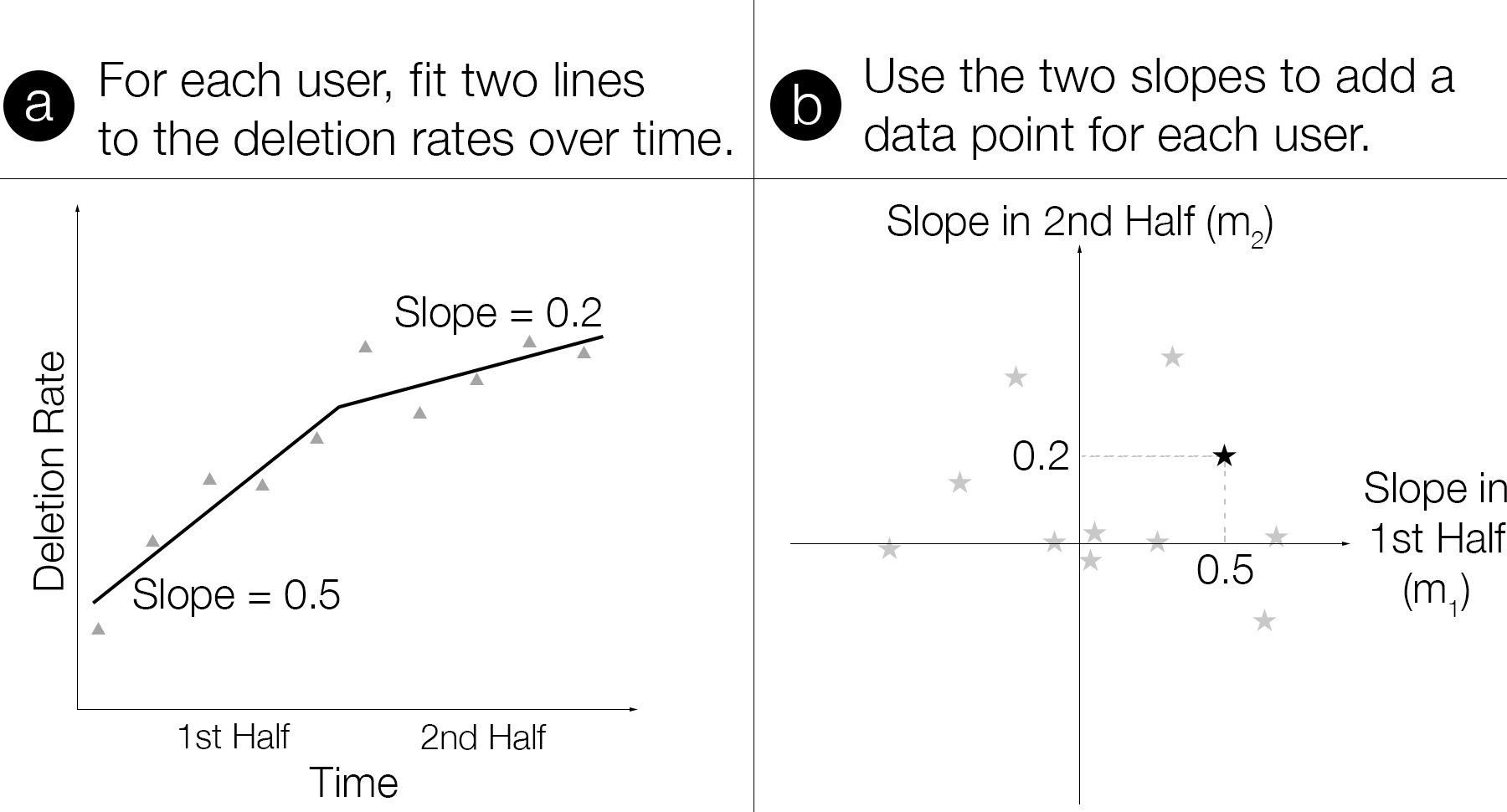}
    \caption{(a) To compute the two-phase trend for a user, two lines are fit to the deletion rates over time, one for the first half of a user's life, and one for the second half. (b) The slopes of these two lines can then be subsequently plotted and compared to that of other users to see how users ``improve'' or ``worsen'' over time.}
    \label{fig:twophase_explanation}
\end{figure}

Motivated by our observations of antisocial behavior over time, we can start to identify types of users based on their behavior in different stages of their life in a community.

\subsection{Users with Differing Deletion Rates}
\label{sec:persistenttrolls}
As Figure \ref{fig:plot_deletion_prop} indicates, the distribution of FBUs based on the proportion of their posts deleted is bimodal.
This suggests that there are roughly two populations of users: FBUs who have a high proportion of posts deleted (\emph{Hi-FBUs}), and FBUs who have a much lower proportion of posts deleted (\emph{Lo-FBUs}).
Here, we define Hi-FBUs as FBUs with a proportion of deleted posts above 0.5, and Lo-FBUs as those with a proportion below 0.5.
Across all communities, the number of users in each population is split fairly equally between the two groups.

Hi-FBUs exhibit characteristics more strongly associated with antisocial behavior: compared to Lo-FBUs, they tend to use language that is less accommodating (\textit{t}>2.3, \textit{p}<0.05, \textit{d}>0.17), receive more replies (\textit{t}>1.8, \textit{p}<0.05, \textit{d}>0.10), and write more posts per thread (\textit{t}>3.9, \textit{p}<10\textsuperscript{-4}, \textit{d}>0.21).
On CNN and Breitbart, Hi-FBUs also swear more (\textit{t}>2.6, \textit{p}<0.01, \textit{d}>0.16).
Unsurprisingly, Hi-FBUs write fewer posts than Lo-FBUs (on average, half as many) (\textit{t}>2.4, \textit{p}<0.01, \textit{d}>0.18) over a shorter period of time before getting banned.

Observing users' post deletion rates over time, for Hi-FBUs, their post deletion rate starts high and remains high (Figure \ref{fig:plot_deleted_bad}).
In contrast, Lo-FBUs tend to have a constant lower post deletion rate, similar to users who were never banned (NBUs), until the second half of their life where it starts to rise significantly.

\begin{figure*}[t]
        \centering
        \begin{subfigure}[b]{0.32\textwidth}
                \includegraphics[width=\textwidth]{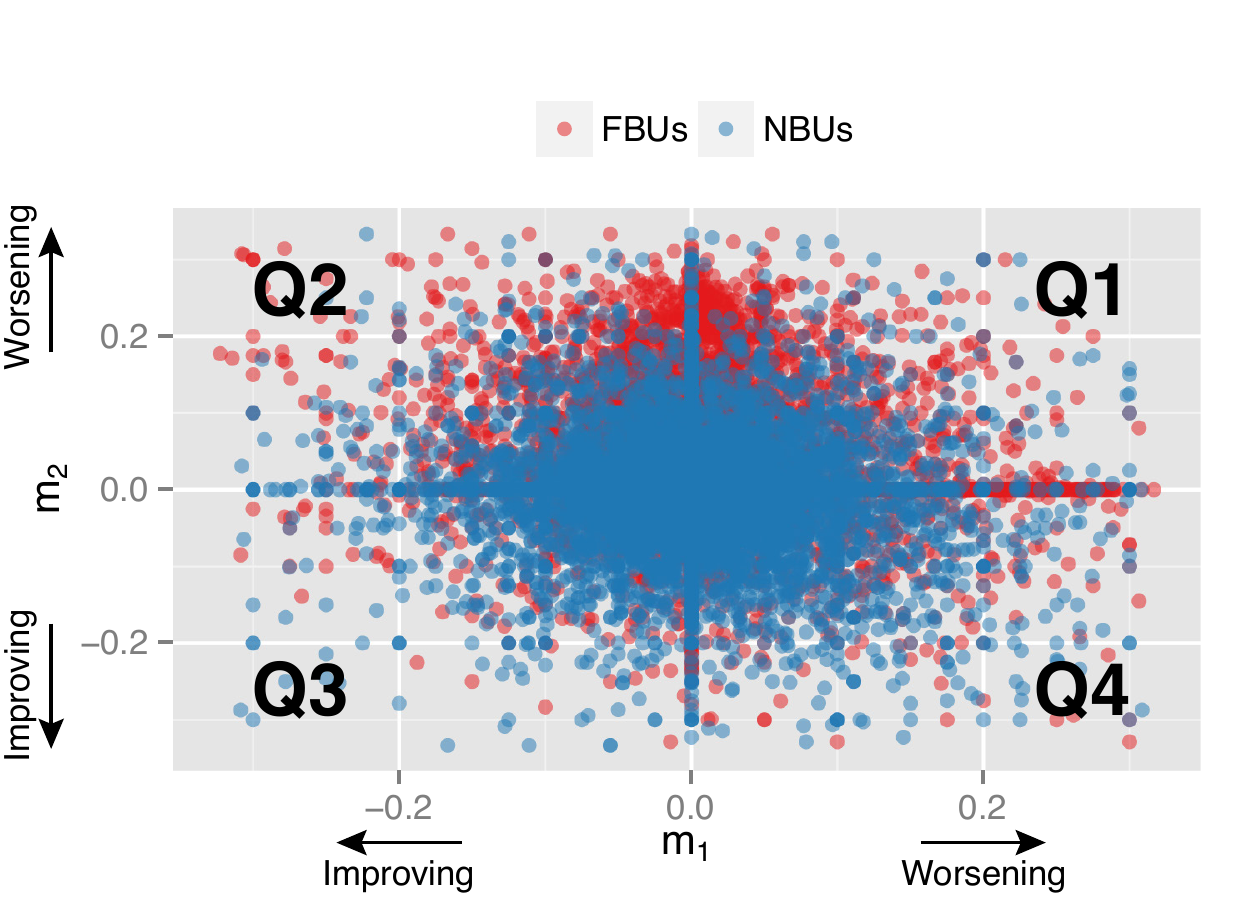}
                \caption{Scatter plot of slopes}
                \label{fig:deleted_2_scatter}
        \end{subfigure}
        \begin{subfigure}[b]{0.32\textwidth}
                \includegraphics[width=\textwidth]{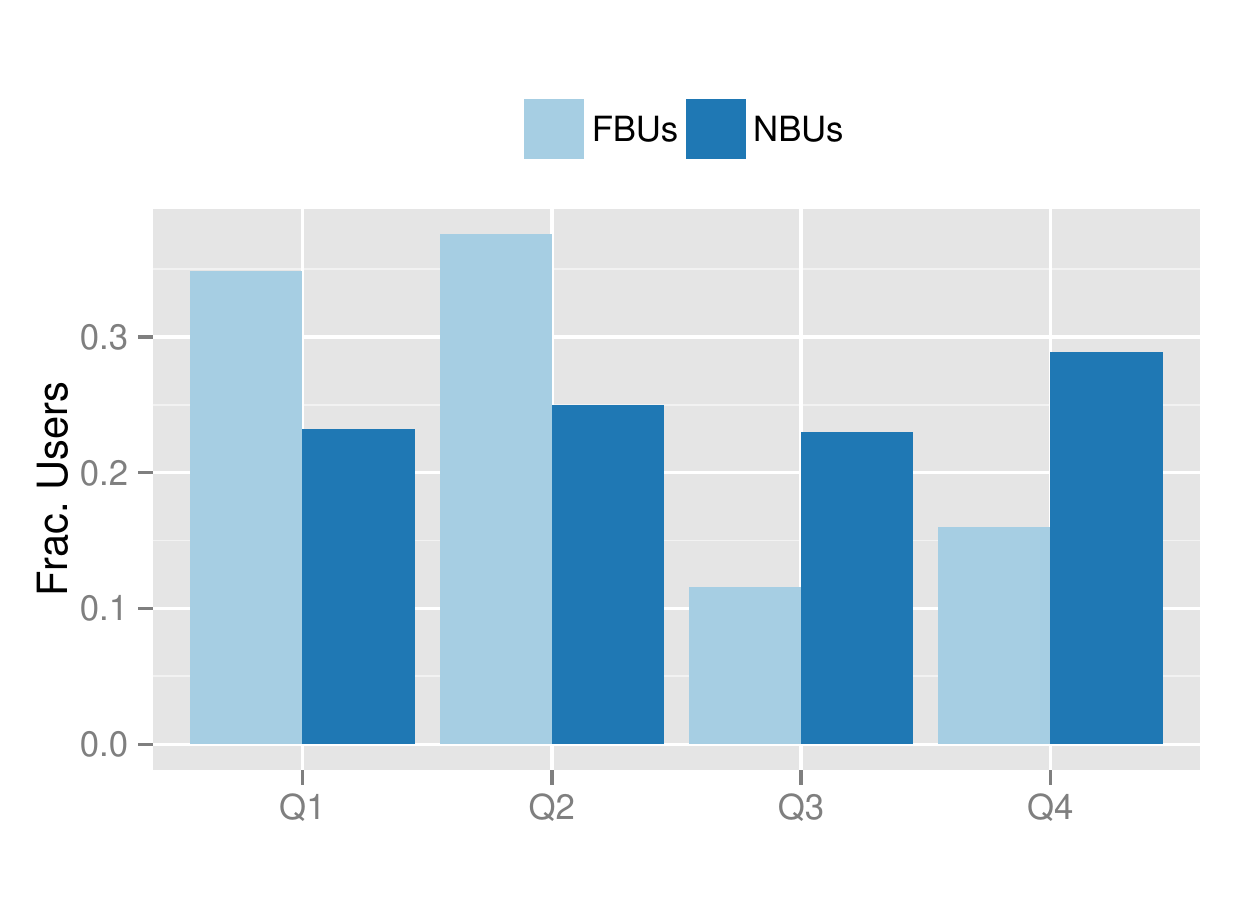}
                \caption{Quadrant dist. for FBUs and NBUs}
                \label{fig:deleted_2_scatter_quadrants}
        \end{subfigure}
        \begin{subfigure}[b]{0.32\textwidth}
                \includegraphics[width=\textwidth]{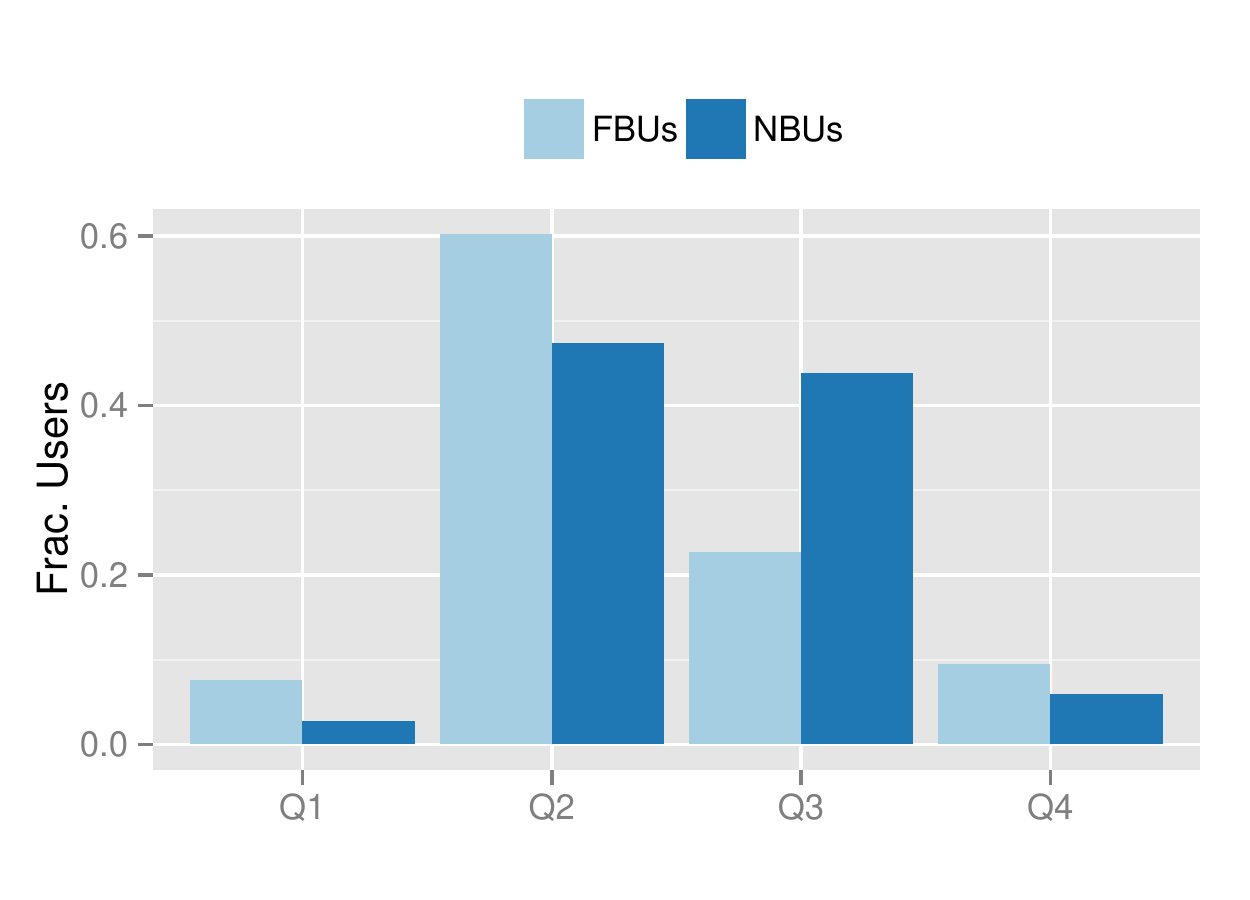}
                \caption{Users with high initial deletion rates}
                \label{fig:deleted_2_scatter_quadrants_redeemed}
        \end{subfigure}%
        \caption{(a) By plotting the slopes in the first ($m_1$) and second ($m_2$) halves of a user's life, we can identify quadrants corresponding to how the deletion rate of a user's posts changes over time. (b) We find that the deletion rate is more likely to continuously increase for FBUs, and decrease for NBUs. (c) A substantial number of users with initially high deletion rates appear to improve over time.}
        \label{fig:twophases_plots}
\end{figure*}

While both types of users are writing worse over time, we hypothesize that this increased deletion rate among Lo-FBUs could stem from a number of factors.
First, Lo-FBUs tend to write less similarly to other users near the end of their lives: the text similarity of these users' posts with other posts in the same thread is significantly lower across their last five posts than their first five (\textit{t}>4.6, \textit{p}<10\textsuperscript{-4}, \textit{d}>0.25) (Figure \ref{fig:plot_lohiprop_cosine}).
Additionally, Lo-FBUs, while initially posting across more threads in the early stages of their life, start to both post more frequently (\textit{t}>2.4, \textit{p}<0.01, \textit{d}>0.04), and in fewer threads (\textit{t}>5.6, \textit{p}<10\textsuperscript{-4}, \textit{d}>0.23) later in their life.
Thus, a combination of a large number of less relevant posts in a short period of time potentially makes them more visible to other members of the community, including community moderators, even if they are writing similarly as before.

\subsection{Antisocial Behavior in Two Phases}
\label{sec:twophases}
Beyond differentiating between Hi-FBUs and Lo-FBUs, our results thus far suggest that we may be able to more generally categorize users based on changes in their behavior over time.
Previously, we showed that changes in both text quality and community perception can result in changes in a user's post deletion rate.
We now attempt to characterize this change over a user's lifetime.

We consider a simple model which splits a user's life into two halves, with the goal of understanding how users may change across them.
We first fit two linear regression lines to a user's post deletion rate over time, one for each half of the user's life (Figure \ref{fig:twophase_explanation}a).
These post deletion rates are obtained by bucketing posts into tenths of a users' life, and computing the proportion of posts deleted in each bucket.
We can then use the the slope in the first half, $m_1$, the slope in the second half, $m_2$, and the intercepts $c_1$ and $c_2$ to quantify how the deletion rate changes over time.
If the slope is positive, the deletion rate is increasing, and thus a user is getting ``worse'' (with respect to activity and bias); if the slope is negative, the deletion rate is decreasing, and a user is getting ``better''.

We can further represent each user as a point $(m_1,m_2)$ to understand user behavior in aggregate (Figure \ref{fig:twophase_explanation}b).
Figure \ref{fig:deleted_2_scatter} shows a scatter plot of the distribution of $m_1$ and $m_2$.
The cloud of red points (FBUs) is shifted upwards relative to that of NBUs, indicating that overall, FBUs have more of their posts deleted over time.
By plotting $m_2$ against $m_1$, we can also observe the relative change in deletion rates over time.
For instance, in Quadrant 1 (Q1 in Figure \ref{fig:deleted_2_scatter}), both $m_1$ and $m_2$ are positive, indicating that the deletion rate for users in that quadrant is increasing in both halves of their life.
In contrast, for users in Q2, the deletion rate decreases, then increases ($m_1$<0, $m_2$>0).
We find that the fraction of users who are getting worse (Q1) is higher for FBUs than NBUs (\textit{p}<0.01) (Figure \ref{fig:deleted_2_scatter_quadrants}). Further, the fraction of users who are improving (Q3) is higher for NBUs than FBUs (\textit{p}<0.01).

Similar to our previous analysis of Hi- and Lo-FBUs, the change in deletion rates can be partially explained by the changes in the text similarity of a post and previous posts in the same thread: in all quadrants, when the text similarity increases, the deletion rate decreases, and vice versa.

\xhdr{Characterizing users who were not banned}
Throughout this paper, we use ground truth data (\ie whether a user has been banned) to characterize antisocial users.
However, moderators may not catch all users engaging in antisocial behavior.
For instance, some NBUs may have been initially deviant, but improved over time.
Here, we identify users who have an initially high proportion of deleted posts ($\ge$ 0.5), and compare the populations of users who were eventually banned (FBUs) and those who were not (NBUs).

As shown in Figure \ref{fig:deleted_2_scatter_quadrants_redeemed}, overall, a smaller proportion of NBUs get consistently worse (Q1)\footnote{The relatively low proportion of users in Q1 overall stems from the fact since the deletion rate is initially high, few users actually get even worse over time.}, and a larger proportion of NBUs do improve (Q3) (\textit{p}<0.01).
We also find a substantial number of FBUs in Q3, implying that although these users are improving in a sense, they still get banned.
On the other hand, the high proportion of NBUs in Q2 suggests that many users who should be banned are in fact not.
Nonetheless, this approach does not identify all users that exhibit such behavior (e.g., those who have middling post deletion rates but later become model community members).

\section{Identifying Antisocial Users}
\label{sec:prediction}

In the communities we studied, users who are subsequently banned from a community (FBUs) tend to live for a long time before actually getting banned, suggesting that these communities tend to respond slowly to toxic users.
On CNN, FBUs write an average of 264 posts (over 42 days), with 124 posts deleted before they are finally banned.
Thus, we turn our attention to building tools that could allow for automatic, early identification of users who are likely to be banned in the future.
With only a user's first ten posts, we find that we can accurately differentiate FBUs from NBUs.
By finding these users more quickly, community moderators may be able to more effectively police these communities.

\begin{table}[tb]
\small
\centering
\ra{1.3}
\begin{tabular*}{\columnwidth}{lp{6cm}}\toprule
  \textbf{Feature Set} & \textbf{Features} \\
  \hline
  Post (20) & number of words, readability metrics (e.g., ARI), LIWC features (e.g., affective) \\ 
  Activity (6) & posts per day, posts per thread, largest number of posts in one thread, fraction of posts that are replies, votes given to other users per post written, proportion of up-votes given to other users \\
  Community (4) & votes received per post, fraction of up-votes receieved, fraction of posts reported, number of replies per post \\
  Moderator (5) & fraction of posts deleted, slope and intercept of linear regression lines (i.e., $m_1$,$m_2$,$c_1$,$c_2$) \\
\bottomrule
\end{tabular*}
\caption{We considered four categories of features, which in order correspond to having increasingly more information about a user's behavior in a community.}
\label{tab:feature_list}
\end{table}

\subsection{Factors that help identify antisocial users}
Using the observations and insights from previous sections, we begin by designing features that can help identify antisocial behavior in a community early on.
We group them into four categories (Table \ref{tab:feature_list}):

\xhdr{Post features}
A natural predictor of a post's undesirability the content of the post itself.
We previously found that posts written by FBUs are less readable (and thus include readability metrics), and differ in affective content such as swearing \cite{wang2010got}.
In our initial analysis, features such as capitalization and punctuation \cite{adler2011wikipedia,javanmardi2011vandalism} were not as informative; sentiment classifiers did not provide significant performance benefits above the affective categories from LIWC.

\xhdr{Activity features}
In addition to writing differently, FBUs also differ from NBUs in their activity.
For instance, FBUs tended to spend more time in individual threads than NBUs.
Prior work also identified post frequency as a signal of a low quality discussion \cite{diakopoulos2011towards}.
Thus, we include features such as the proportion of posts that are replies and the maximum number of posts in a single thread, in addition to other features such as posts per day or votes given to other users.

\xhdr{Community features}
We also considered the different mechanisms in which other members of a community interact with users and their posts.
For instance, a low proportion of up-votes received has been shown to be perceived negatively by users \cite{cheng2014community}.
As we only examine the first ten posts of a user, indicators of reputation, while useful in other settings \cite{adler2011wikipedia}, are unlikely to be informative here.

\xhdr{Moderator features}
Moderator features (\ie~features relating to post deletion) constitute the strongest signals of deletion, as community moderators are responsible for both deleting posts and banning users.
In addition to the proportion of posts deleted so far, we include the slopes and intercepts of the deletion rate over the first ten posts.
We also consider the proportion of deleted posts as a strong baseline to improve upon.
Though moderator features alone outperform any other feature set, we can still achieve substantially greater performance when all feature sets are aggregated.

\begin{table}[tb]
\small
\centering
\ra{1.3}
\begin{tabular*}{\columnwidth}{@{\extracolsep{\fill}}llll}\toprule
   & \textbf{CNN} & \textbf{IGN} & \textbf{Breitbart} \\
  \hline
  Bag-of-words & 0.70 & 0.72 & 0.65 \\
  \hline
  Prop. Deleted Posts & 0.74 & 0.72 & 0.72 \\
  \hline
  Post & 0.62 & 0.67 & 0.58 \\
  + Activity & 0.73 \deem{(0.66)} & 0.74 \deem{(0.65)} & 0.66 \deem{(0.64)} \\
  + Community & 0.83 \deem{(0.75)} & 0.79 \deem{(0.72)} & 0.75 \deem{(0.69)} \\
  + Moderator & 0.84 \deem{(0.75)} & 0.83 \deem{(0.73)} & 0.78 \deem{(0.72)} \\
\bottomrule
\end{tabular*}
\caption{A classifier that uses post, activity, community, and moderator features is able to accurately predict whether a user will be subsequently banned, and performs substantially better than a bag-of-words classifier or relying on a moderator to manually identify posts that should be deleted. Shown are the performance improvements from incrementally adding these features, with AUC reported. Individual feature set performance is in parentheses.}
\label{tab:performance}
\end{table}

\subsection{Predicting antisocial behavior}
With these features in mind, we consider a prediction task where we observe a user's first ten posts, and predict whether this user will eventually get banned.
Here, we perform the above task on a balanced dataset of FBUs and NBUs ($N$=18758 for CNN, 1164 for IGN, 1138 for Breitbart).
In other words, exactly half of users are eventually banned, and random guessing achieves a classification accuracy of 50\%.
Using a random forest classifier, we performed 10-fold cross validation and report the area under the ROC curve (AUC). All features were standardized.

As shown in Table \ref{tab:performance}, we can accurately predict whether a user will be subsequently banned with a mean AUC of 0.82 (mean accuracy=0.74, mean F1=0.71).
A logistic regression classifier gives empirically similar results.
The classifier remains robust even in the absence of moderator features (mean AUC=0.79), which may be more difficult to obtain as they essentially require manual labeling of posts.
If we rely on moderators to identify deleted posts, and only use the proportion of posts deleted so far as our predictor, we obtain a mean AUC of 0.73.
A baseline bag-of-words model that used logistic regression trained on bigrams performs reasonably well (mean AUC=0.69), but is less generalizable across communities as we later show.
Instead comparing FBUs with all other users, as opposed to similarly active users that were never banned (NBUs), results in slightly better performance (mean AUC=0.87).

To understand the relative importance of these features, we computed the classification performance of each individual feature using logistic regression.
Unsurprisingly, moderator features are the strongest predictors of being subsequently banned (individual feature set AUC=0.75 for CNN), with the most performant feature being the proportion of posts deleted (0.73).
Community features were next strongest (0.75), with a lower proportion of up-votes received (0.67) and a higher number of reported posts (0.66) both indicators of antisocial behavior. The text similarity of a post with previous posts in a thread, while correlated with post deletion, does not improve classifier performance.
Activty features follow (0.66), with the number of posts per day the most indicative (0.64) of being subsequently banned.
Post features were collectively the weakest predictors of whether a user will ultimately get banned (0.62).
Future work could involve identifying better textual features (\eg phrase structure), and take the context of the post (\ie~the surrounding posts) into account.

\begin{figure}[t]
        \centering
        \begin{subfigure}[b]{0.48\columnwidth}
                \includegraphics[width=\textwidth]{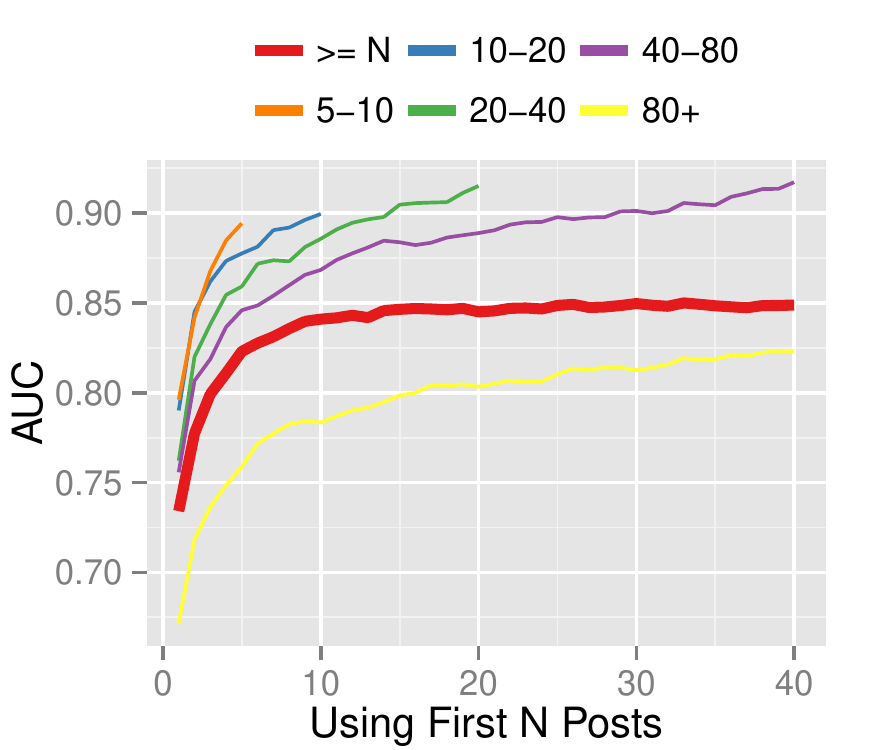}
                \caption{Performance against \# posts}
                \label{fig:performance_change}
        \end{subfigure}
        \begin{subfigure}[b]{0.48\columnwidth}
                \includegraphics[width=\textwidth]{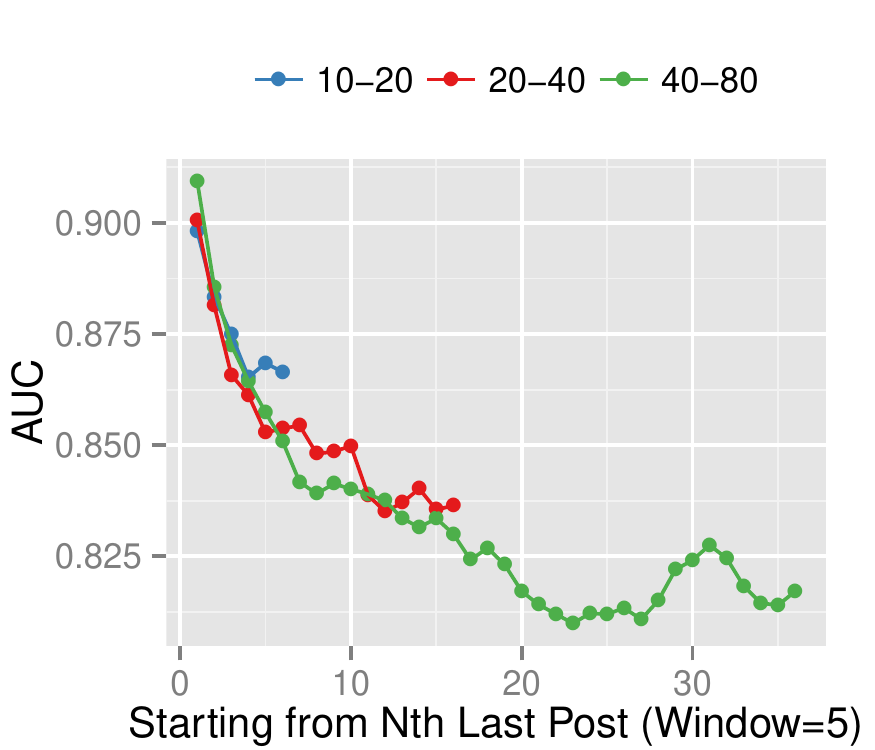}
                \caption{Performance against time}
                \label{fig:performance_sliding}
        \end{subfigure}
        \caption{(a) Prediction performance increases as more posts are observed. Additionally, users who live longer are more difficult to identify early on. (b) Using a sliding window of five posts, performance decreases with increasing temporal distance from the time of deletion.}
        \label{fig:performance}
\end{figure}

\xhdrr{How does prediction performance change with the number of posts observed?}
If the classifier only has access to the first five posts, it can still predict whether a user will get banned with an AUC of 0.80 (across all communities). More generally, performance seems to peak near ten posts (Figure \ref{fig:performance_change}). The same figure also shows how classifier performance changes for users with different post counts: the more posts a user eventually makes, the more difficult it is to predict whether they will get eventually banned later on.

\xhdrr{How does prediction performance change with ``distance'' from when a user gets banned?}
Instead of looking at a user's initial posts, we now consider sliding windows five posts in width, starting from the last five posts of a user, to understand if posts made further in the past are as effective as predicting whether a user will get banned.
As Figure \ref{fig:performance_sliding} shows, it becomes increasingly difficult to predict whether a user will subsequently get banned the further in time the examined posts are from when the user gets banned.
This suggests that changes in both user or community behavior do occur leading up to a ban.

\xhdrr{How does the classifier perform on different types of users?}
We previously identified two types of FBUs: those with high deletion rates (Hi-FBUs), and those with low deletion rates (Lo-FBUs).
Overall, the classifier identifies Hi-FBUs (mean recall=0.99) more reliably than Lo-FBUs (mean recall=0.41).
As Hi-FBUs exhibit high deletion rates from the beginning of their life while Lo-FBUs do not, features such as the proportion of deleted posts are highly informative in the former case but not the latter.
Further, Lo-FBUs write more posts in total than Hi-FBUs, so overall we are examining a smaller fraction of a Hi-FBU's total life in a community.
Still, by assigning higher weights to instances of Lo-FBUs during training, we can maintain a similar AUC, while increasing recall of Lo-FBUs to 0.57 (but decreasing that of Hi-Del users to 0.97).
If we are only interested differentiating Lo-FBUs from other users, a classifier trained on the same features obtains a mean AUC of 0.79 across all communities (recall of Lo-FBUs=0.83). In this case, we may use other mechanisms to identify Hi-FBUs separately.

\begin{table}[tb]
\small
\centering
\ra{1.3}
\begin{tabularx}{\columnwidth}{c>{\raggedleft}X|XXX}\toprule
  & & \multicolumn{3}{c}{Trained on} \\
  & & \textbf{CNN} & \textbf{IGN} & \textbf{Breitbart} \\
  \hline
  {\multirow{3}{*}{\rotatebox[origin=c]{90}{Tested on}}} & \textbf{CNN} & 0.84 & 0.74 & 0.76 \\
  & \textbf{IGN} & 0.69 & 0.83 & 0.74 \\
  & \textbf{Breitbart} & 0.74 & 0.75 & 0.78 \\
\bottomrule
\end{tabularx}
\caption{Cross-domain classifier performance (AUC) is relatively high, suggesting that these learned models generalize to multiple communities.}
\label{tab:crossdomain}
\end{table}

\xhdrr{How generalizable are these classifiers?}
Using a model that uses all four feature sets, we find that cross-domain performance is high (mean AUC=0.74) relative to within-domain performance (Table \ref{tab:crossdomain}), suggesting not only the applicability of these features, but also the generalizability of models learned on single communities.
Most striking is that a classifier trained on the Breitbart community only performs slightly worse if tested on CNN (0.76) or IGN (0.74) than on Breitbart itself (0.78).
We further note that bag-of-words classifiers do not generalize as well to other communities (mean AUC=0.58).

\section{Discussion \& Conclusion}
\label{sec:discussion}

This paper presents a data-driven study of antisocial behavior in online discussion communities by analyzing users that are eventually banned from a community.
This leads to a characterization of antisocial users and to an investigation of the evolution of their behavior and of community response: users that will eventually be banned not only write worse posts over time, but the community becomes less tolerant of them.
Next, it proposes a typology of antisocial users based on post deletion rates. Finally, introduces a system for identifying undesired users early on in their community life.

By using explicit signals of undesirability (i.e., permanent banning), we are able to study users engaged in a wide variety of antisocial behavior.
While scalable, our apporach has several limitations.  A more fine-grained labeling of users (perhaps through crowdsourcing), may reveal a greater range of behavior. Similarly, covert instances of antisocial behavior (e.g., through deception) might be significantly different than overt inflamatory behavior \cite{hardaker2013uh}; some users might surreptitiously instigate arguments, while maintaining a normal appearance.

Further, a better analysis of the content of posts, and of the relation between the posts in a thread, has the potential to reveal patterns associated with discussions stired by antisocial users, e.g., if trolls purposefully ask overly naive questions or state contrary viewpoints \cite{hardaker2010trolling}.

Another future direction is developing a richer taxonomy of antisocial behavior.
Deeper analyses of the differences among groups of users may reveal subtleties in how antisocial users behave (e.g., if different users favor different types of arguments).
To better characterize the different stages of a user's life, future work could explore different models beyond piecewise linear models (\eg locally weighted models).
We also restricted our analysis to users who were permanently banned; studying users who were only temporarily banned may shed more light how some users may redeem themselves in a community.
Additionally, antisocial behavior may also differ in communities other than the ones we studied (e.g., small special-interest communities).

Understanding how antisocial users may steer individual discussions can help us better quantify their influence on other users.
Our initial explorations suggest that FBUs cause discussions to veer off-topic: replies to FBUs were significantly less similar to preceding posts in a thread than replies to NBUs.
One could also investigate the effects of having multiple antisocial users participate in a discussion.

While we present effective mechanisms for identifying and potentially weeding antisocial users out of a community, taking extreme action against small infractions can exacerbate antisocial behavior (\eg unfairness can cause users to write worse).
Though average classifier precision is relatively high (0.80), one in five users identified as antisocial are nonetheless misclassified.
Whereas trading off overall performance for higher precision and have a human moderator approve any bans is one way to avoid incorrectly blocking innocent users, a better response may instead involve giving antisocial users a chance to redeem themselves.

\xhdrr{Acknowledgements}
We thank Disqus for the data used in our experiments and our reviewers for their helpful comments.
This work was supported by a Stanford Graduate Fellowship and a Google Faculty Research Award.

\bibliographystyle{aaai}

\end{document}